\documentclass[aps,showpacs,letterpaper]{revtex4}
\usepackage{graphicx}
\usepackage{hyperref}
\begin{document}

\title{An extended solution space for Chern-Simons gravity: the slowly rotating Kerr black hole}
\author{Mauro Cambiaso$^{1,2}$ and Luis F. Urrutia$^{1}$}
\affiliation{$^{1}$ Instituto de Ciencias Nucleares,
Universidad Nacional Aut{\'o}noma de M{\'e}xico,
A. Postal 70-543, 04510 M{\'e}xico D.F.}
\affiliation{$^{2}$ Departamento
de Ciencias Fisicas, Universidad Andres Bello, Republica 220, Santiago, Chile}

\begin{abstract}
In the Einstein-Cartan formulation, an iterative procedure to find
solutions in  non-dynamical Chern-Simons (CS) gravity
in vacuum is proposed.  The iterations,
in powers of a  small parameter $\beta$ which codifies  the CS coupling, start from an arbitrary
torsionless solution of Einstein equations.
With Schwarzschild  as the zeroth-order choice, we derive a second-order
differential equation for the $\mathcal{O}(\beta)$ corrections to the metric,
for an arbitrary  zeroth-order embedding parameter. In particular, 
the slowly rotating Kerr metric is an $\mathcal{O}(\beta)$ solution in either the canonical 
or the axial embeddings.
\end{abstract}

\maketitle

The Chern-Simons  modification of general relativity
(GR), which we will refer to as CS-EH, was introduced
in \cite{Jackiw:2003pm} within  the framework of the
Einstein-Hilbert formulation of GR and has been the
subject of numerous investigations in the literature,
e.g., \cite{Alexander:2009tp} and references therein.
One of the main topics has been the
search for black hole solutions in the modified theory.
The Schwarzschild solution extends to the so called
canonical CS-EH \cite{Jackiw:2003pm} and some slowly
rotating black hole solutions, which do not include Kerr,
have also been found
\cite{Konno:2007ze,Grumiller:2007rv, Konno:2009kg, Yunes:2009hc}.
Nevertheless,  it has remained  unsolved  the  important
open question to determine under which conditions, if any,
the Kerr black hole arises as a solution of a CS modified
gravity. It is
known that the Kerr metric cannot be a solution of CS-EH because
the Pontryagin constraint is not satisfied, so our finding is
based in an alternative way of implementing the CS modification
to gravity. In fact, the slowly rotating Kerr solution is found
by studying non-dynamical CS modified gravity in vacuum
in the framework of the Einstein-Cartan
(EC) formulation. As we will see, both descriptions are not equivalent
even in the zero torsion limit, resulting in a different space
of solutions for both theories.

We start from the action proposed in \cite{Cantcheff:2008qn},
whose dynamical variables
are the tetrad $\mathbf{e}^{b}$ and the spin connection
$\mathbf{\omega }_{\;\;b}^{c}$ one-forms
\begin{equation}
S[e^{c},\omega^{a}{}_b]=\frac{\kappa }{2}\!\int \!\!
\left( \mathbf{R}%
_{ab}\wedge ^{\ast }\!(\mathbf{e}^{a}\wedge \mathbf{e}^{b})+\beta
\vartheta \;\mathbf{R}^{a}{}_b \wedge \mathbf{R}^{b}{}_a \right) \!.
\label{ACTION}
\end{equation}%
The Riemann $\mathbf{R}_{ab}$ and torsion $\mathbf{T}^{a}$
two-forms are defined as usual
\begin{equation}
\mathbf{R}^{a}{}_b=\mathbf{d \wedge \omega}^{a}{}_b+\mathbf{\omega }
^{a}{}_c \wedge \mathbf{\omega}^{c}{}_b\,,\;\;\;\quad \mathbf{T}%
^{a}=\mathbf{d}\wedge \mathbf{e}^{a}+\mathbf{\omega}^{a}{}_b\wedge
\mathbf{e}^{b}\,.  \label{DEF R and T}
\end{equation}
We refer to this approach as CS-EC, which has also been
recently studied in
\cite{Alexander:2008wi,Kazmierczak:2008iw,Ertem:2009ur}.
In the action (\ref{ACTION}), $\vartheta $ is the so called embedding
parameter and it will be considered as non-dynamical in this work. We take $\beta $ as a smallness dimensionless
parameter, which serves as a bookkeeping device, so that
$[\vartheta ]=L^{2}$ in accordance with $[\kappa ]=L^{-2}$.

Focusing on perturbative solutions to the equations of motion,
we propose a general algorithm based upon an iterative expansion in $\beta$ for all the
variables involved.  This algorithm is an extension of the method
considered in \cite{Cantcheff:2008qn},
allowing us to enlarge the space of solutions to include
a slowly rotating Kerr metric as a first-order perturbation
of the Schwarzschild solution. In \cite{Cantcheff:2008qn}, the
author considers the perturbative effect of the CS term
only in the spin connection, which is then solved for
iteratively in terms of $\mathbf{e}^a$ and $\vartheta$.
However, this procedure, even to first order in $\beta$,
produces a third-order partial differential equation for
the tetrad, and the situation gets even worse to
increasing orders. Our proposal is to extend the perturbative expansion
to all the basic variables, including the embedding
parameter, by writing
\begin{equation}
e_{\mu}{}^{c}= e^{(0)}_\mu{}^c +\beta \, e^{(1)}_\mu{}^c
+\dots \,,\;\;\quad \omega _\mu{}^a{}_b
=\;\omega _{\;\;\;\mu \;\;\;b}^{(0)\;\;a}+\beta \,\omega
_{\;\;\;\mu \;\;\;b}^{(1)\;\;a}+\dots \,,\;\;\;\quad
\vartheta (x)=\vartheta
^{(0)}(x)+\beta \,\vartheta ^{(1)}(x)+\dots \,,
\label{EXP}
\end{equation}
together with analogous expression for all functions
of them. A first advantage of the proposed expansion
is that we do get a second-order differential equation
for the tetrad at each order in $\beta $, as we
explicitly show  in the following. Furthermore,
this construction not only allows to recover the
noteworthy persistence to first order 
in CS-EC of some classical GR solutions, but also allows us to find new solutions,
where the CS term acts as a perturbation of a given
GR solution. In fact, our iterative procedure starts from
a zeroth-order approximation which  might be taken as
any vacuum solution of Einstein equations with zero torsion and provides a systematic
way of constructing the higher-order corrections defined in Eq. (\ref{EXP}).
Since the  approach is non-dynamical in $\vartheta$, consistency is achieved
by further imposing the conservation laws together with the
Bianchi identities upon the equations of motion obtained by
extremizing the action (\ref{ACTION}) with respect to the gravitational
variables only. We consider the implementation of the CS
coupling in terms of the SO(1,3) gauge connection of GR,
which  provides a source of torsion, even in the
absence of matter \cite{REVTORSION,Shapiro:2001rz}. Normally,
torsion enters via the coupling of gravity to fermions and
its presence in nature is severely constrained by observations
\cite{Kostelecky:2007kx}, which provides additional
motivation  for the iterative approach.

Regarding the proposed action (\ref{ACTION}), we observe that
there are additional topological invariants which
include torsion, such as the Nieh-Yan and Holst terms
which may naturally be added to it. The
first  case has been considered in \cite{Ertem:2009ur,Mielke}.
Nevertheless, such interesting additions are  out of
the scope of the present work, which mainly aims to study the possibility of
enlarging the space of solutions in
CS extended gravity. The dynamics of CS gravitational terms, in the context of string theory, has been
previously studied in relation to the search for axionic hair in black holes
\cite{AXIONHAIR}. Instead of that in Eq.\,(\ref{ACTION}), the
interaction considered in \cite{AXIONHAIR} was
$S_{int}[B_{\mu \nu },g_{\alpha \beta }]\sim \int {\mathbf H} \wedge{}^{\ast}{\mathbf H}$,
with  ${\mathbf H}={\mathbf d} \wedge {\mathbf B}+\gamma {\mathbf C}_{RR},$ where the CS 3-form $%
{\mathbf C}_{RR}$ is such that ${\mathbf d} \wedge {\mathbf C}_{RR}=-\frac{1}{2}{\mathbf R}^\alpha{}_\beta
\wedge {\mathbf R}^{\beta }{}_{\alpha}$ and $B_{\mu\nu}=-B_{\nu\mu}$ is the Kalb-Ramond field.

In order to have a self-contained presentation and also to fix our notation we
give a brief account  of some  properties of the EC formulation. A concise notation and an efficient
way of manipulating the Bianchi identities together with
the equations of motion derived from (\ref{ACTION}), is obtained by  introducing the
exterior covariant derivative $\mathcal{D}$ acting upon
Lorentz-valued tensor $p-$forms
$\mathbf{F}_{\;\;\;\;\;rs....}^{ab....}$, such  that the
following $%
(p+1)-$form is produced
\begin{equation}
\mathcal{D} \wedge \mathbf{F}^{a b \dots}{}_{r s \dots }
=\mathbf{d \wedge F}%
^{ab\dots}{}_{rs \dots}\!+\mathbf{\omega }^{a}{}_m\wedge
\mathbf{F}^{mb \dots}{}_{rs \dots }+\cdots -\,\mathbf{\omega }%
^{m}{}_r\wedge \mathbf{F}^{ab \dots}{}_{ms \dots }-\cdots\,.
\end{equation}
The derivative  $\mathcal{D}$ satisfies the Leibnitz
rule and
$\mathcal{D} \wedge \eta ^{ab}=0,\,\,\mathcal{D} \wedge
\epsilon ^{abcd}=0$. Thus,  the
Bianchi identities read
\begin{equation}
\mathcal{D}\wedge \mathbf{R}^{a}{}_b=0\,,\quad \quad
\mathcal{D} \wedge \mathbf{T}^{a}=%
\mathbf{R}^{a}{}_m \wedge \mathbf{e}^{m}\,.  \label{BI1}
\end{equation}%
The equations of motion from  (\ref{ACTION}) are
\begin{eqnarray}
\delta \mathbf{e}^{c} &:&\;\;\epsilon _{abcd}\,
\mathbf{R}^{ab}(\omega )
\wedge \mathbf{e}^{c}=0\,,  \label{EINSTEIN} \\
\delta \mathbf{\omega }_{\;\;b}^{a} &:&\epsilon _{abqp}\;
\mathbf{e}^{q}\wedge \mathbf{T}^{p}+2\,\beta\,
\mathbf{d}\vartheta
\;\wedge \mathbf{R}_{ab}(\omega )=0\,.  \label{TORSIONEQ}
\end{eqnarray}%
In the last equation we see that the CS term acts
as a source of torsion, even in the absence of matter
spin-currents. From Eq.\,(\ref{TORSIONEQ}) we can solve
for $\mathbf{T}^{p}$, which we defer until we have
rewritten the equations in terms of tensor components
in a coordinate basis, which are ultimately employed
in the calculations. Imposing (\ref{EINSTEIN}) in
(\ref{TORSIONEQ}) leads to the cyclic identity for
the components of the torsion
\begin{equation}
0=\mathbf{e}^{d}\wedge \mathbf{T}^{c}\wedge
\mathbf{e}_{c}\rightarrow
\epsilon_{a}{}^{bcd}\, T^{a}{}_{bc}=0,  \label{CYCLICT}
\end{equation}%
written here in terms of the torsion components in
the tetrad basis:
$\mathbf{T}^{a}=\frac{1}{2}T^{a}{}_{bc} \mathbf{e}^{b}
\wedge \mathbf{e}^{c}$.

Next we study the consistency
of the equations of motion under the Bianchi identities.
Applying $\mathcal{D\;}$ to (\ref{EINSTEIN})
and (\ref{TORSIONEQ}) respectively leads to
\begin{eqnarray}
0&=&\epsilon _{abcd}\;\mathbf{R}^{ab}\wedge \mathbf{T}^{c}\,,
\label{CONSIST1}\\
0&=&\epsilon _{abqp}\,\mathbf{e}^{q}\wedge \mathbf{R}^{p}{}_{m}
\wedge \mathbf{e}^{m},  \label{CONSIST2}
\end{eqnarray}%
where we have used
$\mathbf{T}^{q}\wedge \mathbf{T}^{p}\;=\mathbf{T}%
^{p}\wedge \mathbf{T}^{q}$, the definition of the torsion,
the property $\mathcal{D}\mathbf{d}\vartheta =0$ and
(\ref{BI1}). In our particular case, the consistency  condition (\ref{CONSIST2})
yields  the symmetry of the
components of the Ricci tensor
\begin{equation}
{ R}_{ab}={ R}_{ba},\;\; { R}_{ab}={ R}^{m}{}_{amb}
\;,\;\;\;\mathbf{R}^{a}{}_{b}=%
\frac{1}{2}{ R}^{a}{}_{bmn}\;\mathbf{e}^{m}\wedge
\mathbf{e}^{n}.
\label{CONSIST3}
\end{equation}%
It is convenient to write the field equations and consistency
conditions in terms of tensor components in a coordinate
basis $\mathbf{d}x^{\mu }$, for which the tetrad and
connection one-forms' components are defined as
$
e^{c}=e_{\mu }{}^c \,\mathbf{d}x^{\mu },\,\, \omega
^{a}{}_{b}\;=\omega_\mu{}^a{}_b \,\mathbf{d}%
x^{\mu }\,.
$
The standard relations and interpretations hold:
$
g_{\mu \nu }= e_{\mu}{}^{a}\, \eta _{ab}\, e_{\nu}{}^{b},
\,\,  g^{\mu \nu }g_{\nu \rho }=\delta _{\rho }^{\mu }, \,\,
e_{\gamma}{}^{a}E_{a}{}^{\beta}=\delta _{\gamma }^{\beta }\,\,
{\rm  and} \,\,
E_{b}{}^{\gamma}e_{\gamma }{}^{a}\;=\delta _{b}^{a}\,.
$
In this notation the Einstein equation (\ref{EINSTEIN}) reads
\begin{equation}
{ G}_{\phantom{\mu}\nu }^{\mu }=0, \quad \rightarrow
\quad { R}=0 \quad \rightarrow \quad
{ R}_{\phantom{\mu}\nu}^{\mu }=0,.  \label{EINSTCOORD}
\end{equation}%
The dual of the 4-form
$\epsilon _{abqp}\mathcal{D}\wedge (\mathbf{R}^{ab}(\omega
))\wedge \mathbf{e}^{c}$, which is zero in virtue of the
first Bianchi identity in Eq.(\ref{BI1}), leads to
\begin{equation}
{ \nabla} _{\alpha }{ G}^{\alpha}{}_{\psi }-T^{\theta
}{}_{\psi \beta } { R}^{\beta}{}_\theta -\frac{1}{2}T^{\beta
}{}_{\alpha \theta } { R}^{\alpha \theta}{}_{\beta \psi }=0,
\label{CONSIST1 AS KOSTELECKY}
\end{equation}%
where ${ \nabla}_\alpha$ denotes the torsionful
covariant derivative. Since
${ R}_{\phantom{\mu}\nu}^{\mu }=0$, consistency
with (\ref{EINSTCOORD}) demands
\begin{equation}
T^{\beta}{}_{\alpha \theta }
{ R}^{\alpha \theta}{}_{\beta \psi }=0\,.
\label{CONSIST4}
\end{equation}%
This condition is the same as the one obtained in
(\ref{CONSIST1}). Even though the result is an implicit highly non-linear
relation for the connection, Eq. (\ref{TORSIONEQ}) can
be solved for $T^{\sigma}{}_{\alpha \beta }$ in terms
of the full curvature yielding
\begin{equation}
T^{\sigma}{}_{\alpha \beta }=\frac{\beta }{2 e}
\left(\partial _{\mu }\vartheta \right)%
\left[ 2\epsilon ^{\mu \nu \rho \sigma }
{ R}_{\alpha \beta \nu \rho }+\left(
\delta _{\alpha }^{\sigma }\epsilon ^{\mu \nu \rho \omega }
{R}_{\beta \nu \rho
\omega }-\delta _{\beta }^{\sigma }\epsilon ^{\mu \nu \rho \omega }
{R}_{\alpha
\nu \rho \omega }\right) \right], \quad
e=\mathrm{det}(e_\mu{}^a).
\label{TEXACT}
\end{equation}
Here $\epsilon ^{\mu \nu \rho \omega }=0, \pm 1$ is
the Levi-Civit\`a symbol. This equation will provide
the starting point to implement our iterative algorithm.

Before presenting some details of our proposal we  comment upon the
relationship  between the CS-EH and CS-EC formulations.
In the original CS-EH approach the connection is
\textit{a priori} taken to be the torsion-free
Christoffel connection and so are all the
geometrical quantities that depend on it. The CS term
is then a functional of the metric. In the CS-EC case,
the connection is determined from the field equations,
where (\ref{TORSIONEQ}) reveals that the CS term is the
source of non-vanishing torsion. That both approaches
are inequivalent was already shown in Ref.
\cite{Cantcheff:2008qn} and can be seen from the
following argument. In \cite{Jackiw:2003pm} it was
shown that the Schwarzschild metric (written in the
standard form using  spherical coordinates
$t,r,\theta, \phi$ \cite{SCHWAR}) is a solution to
full CS-EH in the canonical embedding, where
$\partial_\mu \vartheta \sim (1/{m})\delta_{\mu t}$
with $m$ a constant. In the CS-EC case one can
easily verify that the Schwarzschild solution (which is
torsionless) is inconsistent with the full expression
(\ref{TEXACT}). Since the corresponding Riemann tensor
satisfies the cyclic Bianchi identity, the RHS of
(\ref{TEXACT}) reduces only to the first term. Also, the
non-zero components of the Schwarzschild Riemann tensor
have the general form $%
R_{\alpha\beta\alpha\beta}$ (no sum over $\alpha$ or
$\beta$), in such a way that it is enough to consider, for
example, $T^r{}_{\theta\phi} \sim
\epsilon^{tr\theta\phi}R_{\theta\phi\theta\phi}$. This
relation is inconsistent with zero torsion because
$R_{\theta\phi\theta\phi}$ is different from zero. In
others words, the unperturbed Schwarzschild metric
cannot be a solution of the full CS-EC extension, which
is in accordance with the  results found
in \cite{Cantcheff:2008qn}.
In the EC formulation of Ref.\cite{Cantcheff:2008qn}, the usual
second-order formulation of CS-EH is recovered by truncating
the connection to first order and replacing it in the
original action. In this case all the zeroth-order
quantities comprise the usual (torsion-free)
Einstein-Hilbert piece and the first-order ones ultimately
furnish the Cotton-like contribution of the CS-EH formulation.
However, the very fact that by this method the author
of \cite{Cantcheff:2008qn} recovers the
results of the original
CS-EH proposal \cite{Jackiw:2003pm}, implies that the Kerr
solution does not fit in his model either, i.e., any
eventually relevant effects
of torsion are washed away by plugging the first-order
connection in the action. Our findings
reveal that such effects do exist and are interesting, not only for the sake
of generality, but also for the phenomenological consequence that
the proposed CS-EC extension of GR can describe (at least slowly) rotating Kerr black holes.

The general procedure for solving (\ref{EINSTEIN}) and
(\ref{TORSIONEQ}) in the EC formulation  can be viewed
as follows: from the torsion field equation (\ref{TORSIONEQ})
one solves for the connection in terms of the tetrad,
$\omega^a_{\phantom{a}b} =
\omega^a_{\phantom{a}b}(\mathbf{e}^c; \vartheta)$. Then
this connection is used as the one defining the curvature
two-form in the Einstein-Cartan field equation (\ref{EINSTEIN})
and subsequently a set of partial differential
equations for $\mathbf{e}^a$ is obtained. Unfortunately, in the
$\beta \neq 0$ case,  (\ref{TORSIONEQ}) is a highly
non-linear equation for the tetrad, so that this step
cannot be performed easily in general and to make some
progress it is convenient to recur to perturbative
methods in terms of the parameter $\beta$. These methods
have been widely used in the CS-EH formulation.

Now we summarize the main steps leading to the equation for the first-order
correction to the metric according to the expansion in  (\ref{EXP}). We begin with an arbitrary
solution of the Einstein equations in vacuum with zero torsion, which provide the quantities
$e^{(0)}, \omega^{(0)}, T^{(0)}=0$ and $R^{(0)}$. The next step is
the calculation of $T^{(1)}$ according to
Eq. (\ref{TEXACT}), which also incorporates an arbitrary  zeroth-order embedding parameter
$\vartheta^{(0)}$. From $T^{(1)}$ we are able to solve the
linear equation for $\omega^{(1)}$, arising from the
definition of the torsion, in terms of $\partial e^{(1)}$
plus zeroth-order quantities including $\vartheta^{(0)}$ together with the zeroth-order
covariant derivative. The next step is to construct the
first-order Riemann tensor, according to the general
definition (\ref{DEF R and T}). It is clear that this
will introduce an additional derivative to
$\partial e^{(1)}$, such that the final equation
$R^{(1)}_{\mu\nu}=0$ will be of second order in the
unknown $e^{(1)}$, providing the first correction to
the metric $g^{(1)}_{\mu\nu}$. At this point, all first-order 
variables are determined and one could start the
calculation of the second- and higher-order corrections
just by repeating the above scheme.

For an arbitrary embedding parameter $\vartheta^{(0)}$, we next obtain the equation for the first-order perturbation of the metric arising when the
zeroth-order contributions $e^{(0)}, \omega^{(0)}, T^{(0)}=0$ and $R^{(0)}$  correspond to the Schwarzschild metric.
In the framework of CS-EH, perturbations of this geometry
have been studied in \cite{Yunes:2007ss}. Following the
general procedure outlined above, the required first
order torsion is
\begin{equation}
T^{(1)\omega }{}_{\alpha \beta }=\frac{1}{e^{(0)}}
\left(\partial _{\mu}\vartheta ^{(0)}\right)\,\epsilon ^{\mu \nu \rho \omega }
R_{\;\;\;\alpha \beta \nu
\rho }^{(0)}\,.  \label{TUNOF}
\end{equation}%
Inserting the definition of the torsion to first order
in the LHS of (\ref{TUNOF}), we obtain an algebraic
equation for $\omega _{\;\;\beta \nu \mu}^{(1)}$ which
yields
\begin{equation}
\label{eq: omega1}
\omega _{\mu \beta \nu  }^{(1)}=-\frac{1}{2e^{(0)}}\,
\left(\partial _{\lambda}\vartheta ^{(0)}\right)\,
S^{\lambda}{}_{\beta \nu \mu }+\frac{1}{2}\left(
e_{\;\;\;\nu \beta }^{(1)}-e_{\;\;\;\beta \nu }^{(1)}
\right) _{;\mu }+\,\,%
\frac{1}{2}\left( e_{\;\;\;\beta \mu ;\nu }^{(1)}-
e_{\;\;\;\nu \mu ;\beta
}^{(1)}\right) +\,\,\frac{1}{2}\left(
e_{\;\;\;\mu \beta ;\nu
}^{(1)}-e_{\;\;\;\mu \nu ;\beta }^{(1)}\right),
\end{equation}%
where
\begin{equation}
S^{\lambda}{}_{\beta \nu \mu} = \epsilon^{\lambda \delta
\gamma}{}_\mu  R_{\;\;\;\delta \gamma \nu \beta }^{(0)}+
\epsilon^{\lambda \delta \gamma}{}_{\nu
} R_{\;\;\;\delta \gamma \mu \beta }^{(0)}-\epsilon
^{\lambda \delta \gamma}{}_{\beta} R_{\;\;\;\delta \gamma \mu \nu
}^{(0)}\,,\qquad e_{\mu \nu }^{(1)}\equiv
e_{\phantom{(1)a}\nu }^{(1)a}e_{%
\phantom{(0)}a\mu }^{(0)}.
\end{equation}%
Here the semicolon  denotes the zeroth-order covariant
derivative, i.e., with the Schwarzschild connection.
The first-order components of the Riemann tensor are
\begin{equation}
R^{(1)a}{}_{b \mu \nu }=D_{\mu }^{(0)} \omega
_{\;\;\;\nu \;\;\;b}^{(1)\;\;a}
-D_{\nu }^{(0)}\omega
_{\;\;\;\mu \;\;\;b}^{(1)\;\;a}  \label{eq: Riem 1}
\end{equation}%
and the first-order Einstein equations (\ref{EINSTEIN})
reduces to
\begin{equation}
0=R_{\;\;\;b\nu }^{(1)}=E_{\;\;\;\;\;\;a}^{(0)\mu }\;
R_{\;\;\;\;\;\;b\mu \nu}^{(1)a}+E_{\;\;\;\;\;\;a}^{(1)\mu }\;
R_{\;\;\;\;\;\;b\mu \nu }^{(0)a}.
\label{eq: EC1st}
\end{equation}%
Substituting (\ref{eq: omega1}) through (\ref{eq: Riem 1})
in (\ref{eq: EC1st}) and making the further definitions:
\begin{equation}
g_{\mu \nu }^{(1)}\equiv e_{\mu \nu }^{(1)}+e_{\nu \mu }^{(1)}\,,
\qquad g^{(1)}\equiv g^{(0)\mu \nu }g_{\mu \nu }^{(1)}\equiv
2e_{\phantom{(1) \rho}\rho
}^{(1)\rho }\,,
\end{equation}
we arrive to the final expression
\begin{equation}
\frac{\vartheta^{(0)} _{;\lambda ;\alpha }}{e^{(0)}}
\left( \epsilon_{\;\;\;\;\;\mu }^{\lambda \delta \gamma }
R_{\;\;\;\delta \gamma \nu}^{(0)\;\;\;\;\alpha }+
\epsilon _{\;\;\;\;\;\nu }^{\lambda \delta \gamma
}R_{\;\;\;\delta \gamma \mu }^{(0)\;\;\;\;\alpha }
\right) =g_{\nu \alpha
\,;\mu }^{(1)\phantom{;\mu};\alpha }+
g_{\mu \alpha \,;\nu }^{(1)\phantom{;\nu};\alpha }-
g_{\mu \nu \,;\alpha }^{(1)\phantom{ ;\alpha};\alpha
}-g_{\phantom{(1)};\mu ;\nu }^{(1)}\,.  \label{FINALEINSTEIN}
\end{equation}%
which is a second-order differential equation for
the first order correction $g_{\mu \nu}^{(1)}$ to the metric, for an arbitrary embedding parameter $\vartheta^{(0)}$.

In the following we verify that the consistency conditions to
order $\beta $ are satisfied. We recover the cyclic condition on
the torsion (\ref{CYCLICT}) because, starting from
(\ref{TUNOF}), we obtain
$\epsilon ^{\nu \omega \alpha \beta }T_{\omega \alpha
\beta }^{(1)}\sim \left(\partial _{\mu }\vartheta^{(0)}\right)
\left[ g^{(0)\mu \nu
}R^{(0)}-2R^{(0)\mu \nu }\right] =0$ in virtue of the
zeroth-order Einstein equations. The symmetry of
$R_{\mu \nu }^{(1)}$, required by the condition
(\ref{CONSIST3}), is readily verified by realizing that
(\ref{FINALEINSTEIN}), written in the form
$R_{\mu \nu }^{(1)}=0$ is  explicitly symmetric in
$\mu, \nu$. To establish the derivative consistency
condition required by  $G^{(1)\mu }{}_{\nu }=0$,
we start by  showing that the first-order contribution to
(\ref{CONSIST4}):
$T^{(1)\omega}{}_{\alpha \beta }
R^{(0)\alpha \beta }{}_{\omega \psi }$, is indeed zero.
Plugging in expression (\ref{TUNOF}) we have%
\begin{equation}
T^{(1)\omega }{}_{\alpha \beta }
R^{(0)\alpha \beta }{}_{\omega \psi }=\frac{1%
}{e^{(0)}} \left(\partial _{\mu }\vartheta ^{(0)} \right)
\epsilon ^{\mu \nu \rho \omega
}R_{\;\;\;\alpha \beta \nu \rho }^{(0)}
R^{(0)\alpha \beta }{}_{\omega \psi }
\end{equation}
Again, the only non-zero components
$R^{(0)\omega \psi }{}_{\omega \psi }\;$ propagates the
indices of the above in such a way that each term in the
sum will contain
$(\partial_\mu \vartheta) \, \epsilon ^{\mu \omega \psi \omega }$,
which is zero. Satisfying  (\ref{CONSIST4}) to first-order
leads to
$\left( \nabla_{\alpha }G_{\;\;\psi }^{\alpha }\right) ^{(1)}=0$.
From the property $G_{\;\;\;\;\;\;\psi }^{(0)\alpha }=0$,
this condition reduces to the
required $G^{(1)\mu }{}_{\nu ;\mu }=0$ consistency relation.

Let us now show how does the null condition for
the Cotton tensor $C_{\mu\nu}$ required in the
CS-EH formulation, arises from our CS-EC expressions.
We note that our effective Einstein equation to first
order reads
\begin{equation}  \label{Einst eq}
G^{(0)}_{\mu \nu} + \beta \,\, G^{(1)}_{\mu \nu} +
\dots = 0\,,
\end{equation}
which implies that  $G^{(0)}_{\mu \nu}$ and
$G^{(1)}_{\mu \nu}$ separately vanish. Had we not
expanded the tetrad, as it is done in
\cite{Cantcheff:2008qn}, then $g^{(1)}_{\mu \nu} = 0$
and  we can directly read what we have called
$G^{(1)}_{\mu \nu}$ from the LHS of (\ref%
{FINALEINSTEIN}). Yet again, noting that the
zeroth-order Ricci tensor vanishes and using the definition of $C_{\mu\nu}$, we verify
that $G^{(1)}_{\mu \nu}\sim C_{\mu\nu} $. The required null condition
is satisfied in the  canonical choice for the
embedding parameter $\vartheta^{(0)} = t/{m}$,
because then the LHS of (\ref{FINALEINSTEIN})
vanishes identically for the Schwarzschild metric.

Finally we
explore in more detail the case
$\vartheta^{(0)} = t/{m}$, where the  first-order
Einstein equation in CS-EC reads
\begin{equation}  \label{Einstein eq zero Cotton}
0 = g^{(1) \phantom{;\mu} ;\alpha }_{\nu \alpha\,;\mu } +
g^{(1) \phantom{;\nu} ;\alpha}_{\mu \alpha\, ;\nu} -
g^{(1) \phantom{ ;\alpha}
;\alpha}_{\mu \nu\, ;\alpha} -g^{(1)}_{\phantom{(1)};\mu ;\nu}.
\end{equation}
Here we trivially see that $g_{\mu \nu}^{(1)} = 0$,
i.e., no modification to
$g^{(0)}_{\mu \nu} \equiv g^{\mathrm{Schw}}_{\mu \nu}$
whatsoever, satisfies the above equation. Therefore we recover
the result that the Schwarzschild metric remains to be a
solution up to first-order in the CS-EC formulation
\cite{Cantcheff:2008qn}.
Nevertheless, (\ref{Einstein eq zero Cotton}) has certainly
extended the space of solutions of our problem. This
naturally leads to the question whether it is  possible,
for a small CS term, to break the full spherical symmetry
of the Schwarzschild metric,  down to the particular case
of axial symmetry of the Kerr metric.  This motivates to try the
following ansatz
\begin{equation}
g_{\mu \nu}^{(1)} =
f(r, \theta) \,\delta^t_\mu \,\delta^\phi_\nu\,.
\end{equation}
After substituting in (\ref{Einstein eq zero Cotton})
we obtain that only the $t, \phi$ component of these
equations  is not identically zero, yielding
\begin{equation}  \label{field eq f}
0= \frac{-4M f}{r^3} + \frac{2 M f^{\prime\prime}}{r}-
f^{\prime\prime}-
\frac{f^{\ast \ast}}{r^2} + \frac{ f^\ast}{r^2\tan \theta} \,.
\end{equation}
Here $f^{\prime}$ stands for $\partial_r\, f$ and
$f^\ast$ for $\partial_\theta \, f$. Recall that in the
above, zeroth-order covariant derivatives were involved,
which in this case are built from the Schwarzschild
metric, hence the parameter $M$, coming from the typical
$(1-2M/r)$ factor and derivatives of it. Although this
is a non-trivial equation for $f$,
it is straightforward to verify that
$f(r,\theta) = -\frac{2M^2}{r} \sin^2\theta$, solves (\ref{field eq f}).
Choosing
$\beta=a/M$ we can write the complete solution for
the gravitational field
to first-order as
\begin{eqnarray}
g_{\mu \nu} = g^{(0)}_{\mu \nu} +
\beta g^{(1)}_{\mu \nu} + \cdots =
g^{\mathrm{Schw}}_{\mu \nu} - \frac{2Ma}{r}
\sin^2\theta \,\delta^t_\mu \, \delta^\phi_\nu =
g^{\mathrm{slow \,Kerr}}_{\mu \nu}\,,
\end{eqnarray}
where the Kerr metric is written in Boyer-Lindquist
coordinates. Even though the Kerr parameter has
been scaled by hand, the relevant property is the
space dependence of the correction, which matches the
slow rotating Kerr solution. Remarkably, we have also verified
that the LHS of (\ref{FINALEINSTEIN}) is zero, and that the consistency
conditions (\ref{CYCLICT}) through (\ref{CONSIST4}) are satisfied, for an
axial embedding parameter $\vartheta= M\,r\, \cos\theta$,
which produces a spacelike symmetry breaking direction parallel
to the rotation axis of the Kerr metric.
This choice of the embedding parameter suggests that one could interpret the CS modification
to GR as a source of Lorentz invariance violation  \cite{Kostelecky:2007kx}.
Nevertheless, explicit breaking due to  $\vartheta$ as introduced here is shown to be
inconsistent with the EC formulation
\cite{Kostelecky:2003fs}. In this  way, a correct
description of this possible interpretation requires that such
violations arise via a spontaneously broken mechanism.
This implementation, together with some interesting open issues (many of them suggested by the referees)
are out of the scope of the present work and will be dealt with elsewhere \cite{MCLU}.

Summarizing, we have  proposed a systematic modified
iterative scheme which extends the space of solutions of CS gravity in the  EC formulation. As a first
bonus of this extension we have been able to make one step forward in the  longstanding problem of including
Kerr black holes in CS gravity. We have shown that, either with
the canonical or the axial embedding parameters, the
slowly rotating Kerr metric, considered as a
perturbation to the Schwarzschild metric due to the
torsion effects introduced by the CS term, is a solution
of the CS-EC formulation of the theory, up to first-order
in the perturbative expansion.

This work was supported by the grants DGAPA-UNAM-IN 111210
and CONACyT \# 55310. The authors acknowledge useful
discussions with  J. Alfaro and H. Morales-T\'{e}cotl
at the early stages of our investigation in gravitational
Chern-Simons theories. We thank R. Caldwell for providing
Ref. \cite{AXIONHAIR}. { Also, useful comments from R. Lehnert are gratefully acknowledged.}
Many intermediate calculations have been verified with the package GRTENSORII for MAPLE.

\end{document}